# Rydberg Matter clusters of alkali metal atoms: the link between meteoritic matter, polar mesosphere summer echoes (PMSE), sporadic sodium layers, polar mesospheric clouds (PMCs, NLCs), and ion chemistry in the mesosphere

Frans Olofson, Patrik U. Andersson and Leif Holmlid*, Atmospheric Science, Department of Chemistry, University of Gothenburg, SE-412 96 Göteborg, Sweden
*E-mail address: holmlid@chem.gu.se

Abstract

There exists a material which links together the influx of meteoritic matter from interplanetary space, the polar mesosphere summer echoes (PMSE), the sporadic sodium layers, the polar mesospheric clouds (PMCs, NLCs), and the observed ion chemistry in the mesosphere. The evidence in these research fields is here analyzed and found to agree well with the properties of Rydberg Matter (RM). This material has been studied with numerous methods in the laboratory. Alkali atoms, mainly Na, reach the mesosphere in the form of interplanetary (meteoritic, cometary) dust. The planar RM clusters Na$_N$ usually contain $N$ = 19, 37 or 61 atoms, and have the density of air at 90 km altitude where they float. The diameters of the clusters are 10-100 nm from laboratory high precision radio frequency spectroscopic studies. Such experiments show that RM clusters interact strongly with radar frequencies: this explains the radio frequency heating and reflection studies of PMSE layers. The clusters give circular polarized scattering and depolarized scattering of visible light in laboratory experiments: similar effects are observed in light scattering and lidar studies of NLCs. The clusters give the low temperature in the mesosphere by efficient selective radiation at long wavelengths, which is observed in RF emission experiments. The lowest possible stable temperature of the mesopause is calculated for the first time to be 121 K in agreement with measurements, based on the strong optical activity at long wavelengths in RM. Sporadic sodium layers are explained in a unique way as due to shockwaves in the RM layers. Due to the high electronic excitation energy in RM clusters, they induce efficient reactions forming ions of all atoms and molecules in the atmosphere thus providing condensation nuclei for water vapour. This finally gives the visible part of the PMC structure. Especially the sporadic sodium layers and the PMSE give direct evidence for the existence of RM layers at the mesopause. The present contribution fills the gap between and partially replaces the separate theories used to describe the various aspects of these intriguing phenomena.

# 1. Introduction

Polar mesospheric clouds (PMC), also named noctilucent clouds (NLCs), have been observed and studied for a long time in the upper atmosphere, at an altitude of approximately 80-90 km. Most results of the studies indicate that the visible clouds consist of water ice particles [Hervig et al. 2001], but the processes which form the clouds are still quite obscure. While direct sampling methods with rockets have some advantages from the point of the chemical composition of the particles, such methods are invasive and probably destroy particles in high electronically excited states, giving no clues on their existence. The well documented cases where a strong influence is observed of the rocket shock and wake on the ion current signal measured are important in this respect [Horányi et al. 1999, Holzworth et al. 2001]. Thus, optical methods like lidar measurements are better suited for revealing the electronic state of the particles involved in the formation of PMCs. Such methods have provided insight into polarized back-scattering processes, optical absorption processes and possible sizes and forms of particles in the PMCs. The existing evidence points strongly to other types of particles than small water ice particles as responsible for these types of features, with special forms or electronic properties of these particles [Alpers et al. 2001, Baumgarten et al. 2002]. From laboratory studies, it is known that particles (clusters) in highly excited states are easily formed by alkali and hydrogen atoms, even at low temperature and pressure [Holmlid 2002; Badiei and Holmlid 2006]. The large amount of experimental information on such small particles of the Rydberg matter (RM) type now makes it possible to propose that the PMCs exist as a result of the formation of such clusters in the alkali atom containing layer that exists just above the PMC height range. This type of electronically excited but extremely long-lived form of matter which is named RM has been studied in the laboratory with a large number of methods during the last 20 years, following its prediction by Manykin et al. [1980]. The theory of RM has been further developed by Manykin et al. [1992a, b, 2000] and by Holmlid [1998a].

The reasons for the existence of such particles (clusters) in the upper atmosphere are several, and some will be discussed below in the theoretical section. However, it can be stated immediately that it is likely that the surroundings at the altitudes of interest here do not differ strongly from the environment in other closely related atmospheres of planets and comets. In all these cases, the influence of sunlight and of the interplanetary medium will be quite similar. The extended alkali atmospheres at the Moon and Mercury are shown to be due to RM particles in a recent publication [Holmlid 2006b]. These RM particles act as an intermediate storage for the alkali atoms, which are periodically and partially released by sunlight and discharge processes in the atmospheres. Further, the spectroscopic and optical properties of comets are shown to agree well with experimental results from RM, and with theoretical predictions based on the properties of RM. The relatively weak polarization of the scattering of initially randomly polarized sunlight from comets was for example recently explained for the first time by RM clusters [Holmlid 2006a]. The circularly polarized light formed by RM layers at the mesopause is recently proposed to be the source of the so-called homochirality in the biosphere [Holmlid 2009].

Besides the importance in finding a common source i.e. RM of all the various phenomena observed at these altitudes (PMCs, NLCs, PMSE, sporadic sodium layers, ion chemistry etc.), important advances in the specific fields are also described below. They also give evidence for the existence of RM at the mesopause. The absorbance and reflection of RF and radar frequencies (UHF, VHF) in the specific "dusty plasma" of RM in the PMSE effect explains the observations and relaxes the requirement of large, highly charged water ice particles at large altitudes, which are otherwise required to explain this effect. Another item is the explanation of the rapid appearance of the sporadic sodium layers via Coulombic shock waves in the RM layers. The seasonal variation of the PMCs and of the Na and K densities are shown to be coupled to the RM cluster properties. Finally, the first calculation of the lowest possible (mean) temperature of 121 K at the mesopause, independent of any adiabatic effects, is one such important advance: it is based on the special radiative properties of RM and explains the observed super-adiabatic temperature variations..

**2. Rydberg Matter: a summary**

A short summary of the most pertinent properties of the state of matter called RM is given here. For a more complete view of its properties, see other publications from our group. In Holmlid [2002], a short review of the methods used to form RM in the laboratory is given. RM is by definition a well-ordered form of matter. Of course, in most experimental studies only the well-ordered type of material is identified and the non-ideal parts evade observation, so the results give larger weight to the well-ordered parts observed in each experiment. Thus, also less well organized parts of the medium may exist in the form of a non-ideal plasma [Norman 2001; Bonitz et al. 2004; Zelener et al. 2004] or so called dusty plasma (see further below in the section on PMSE).
.
RM composed of alkali atoms is a metallic phase of low density, consisting of small particles in the form of planar clusters with a thickness of one atomic layer under the relevant conditions in the upper atmosphere. These planar clusters have six-fold symmetry in their most stable form, and contain a number of 7, 19, 37, 61 or 91 atoms [Holmlid 1998a; Wang and Holmlid 1998, 2000, 2002]. See Fig. 1. These numbers are the so called magic numbers for close-packed planar clusters, but other forms with two-fold or three-fold symmetry are also observed. An example of a time-of-flight experiment with $K_N$ clusters of different forms is shown in Fig. 2. RM consists in principle of long-lived interacting circular Rydberg states but it is a metallic condensed phase: thus no atomic spectra are observed since the valence electrons are delocalized in the metallic bonding. Circular Rydberg states have large principal quantum number *n* and also large angular momentum quantum number *l = n-1*. (Below, the Bohr description is used with $n_B = l$). Such states are long-lived even as separate atoms, and their interaction gives a very long-lived condensed phase. The RM clusters are most easily formed in desorption of alkali atoms like K from non-metal surfaces. They can contain many different atoms and small molecules [Wang and Holmlid 2002; Badiei and Holmlid 2002c, 2006]. At the conditions of interest here, the RM clusters will be formed mainly of Na and to a lesser extent of K atoms. Also H atoms can form the RM layers. An RM cluster can only be stable for a long time if all species in the cluster have the same value of $n_B$, here called the excitation level of RM. The Rydberg electrons in the clusters are delocalized in a conduction band as in a metal, but retain their large angular momentum. The increased stability of the RM phase relative to separate Rydberg species is due to a quantum mechanical interaction of the exchange-correlation type, similar to the case of ordinary chemical bonding. It is also important to observe that the Born-Oppenheimer approximation is not generally valid for RM, i.e. the electrons at high $n_B$ cannot be considered to move in the field from the ions since their velocities are comparable to those of the ions. Thus, both electrons and core ions are particles that move in their mutual field, which is a system that can not yet be solved by quantum mechanical methods. Further, the spacing between some types of RM electronic states may be as small as $10^{-5}$ cm$^{-1}$ (300 kHz) [Holmlid 2004b], while the rotational spacings for the RM clusters will be many orders of magnitude larger [Holmlid 2007b]. Thus, the inverse Born-Oppenheimer approximation may be more applicable. This implies that the electrons provide the main structure in which the ion cores are moving. Conversely, the ion core motion is influenced directly by the electron motion.

One important point to observe is that RM is a condensed phase, where the valence electrons from the atoms in the RM phase are delocalized. The electrons move in the conduction band of the metallic material. This means that for example the valence electron of the Na atom is not in any atomic state, but the $Na^+$ ion is embedded in an electronic sea containing orbiting electrons which shield the ions from each other, as in an ordinary metal. Thus, no electrons are in atomic states and no atomic spectroscopy can identify Na atoms in the RM phase, in the same way as there is no atomic spectrum of Na visible in the light from a piece of ordinary metallic Na. Thus, spectroscopic methods like sodium lidars can not observe any Na atoms in the RM phase. This is so since the valence electrons, giving the normal atomic resonance lines, are in entirely different electronic states.

The distance between two atoms in RM is $d = 2.9\, n_B^2\, a_0$, where $n_B$ is the principal (Bohr model) quantum number (the excitation level) and $a_0$ is the Bohr radius. The approximate factor 2.9 is found from the quasi-classical calculations of the minimum energy states of RM with electron correlation included [Holmlid 1998a]. High-resolution radio frequency spectroscopy studies of the rotation of RM clusters give this relation with much higher precision, for example as 2.879 ± 0.001 for a typical cluster $K_{19}$ in $n_B = 5$ [Holmlid 2007b]. The diameter of such a common cluster is 15.2 nm (see further Table 1). The bond distances in alkali metal RM clusters in excitation level $n_B = 4 - 7$ have also been measured directly from the kinetic energy release by laser initiated Coulomb explosions in the RM clusters [Badiei and Holmlid 2002c, 2002d, Åkesson et al. 2006]. As seen from Table 1, the molecular clusters forming RM have diameters in the range 10 – 100 nm for 19 – 91 atoms. This is a size range usually considered to be due to particles containing a much larger number of atoms, but due to the large interatomic distances for RM, large particles (clusters) contain rather few atoms.

Quantum mechanical solid state calculations of the electron structure were performed to predict a range of properties of RM by Manykin et al. [1992a, b]. They predicted that no absorption of electromagnetic radiation takes place below a typical cutoff wavelength for each excitation level, depending on the low plasma frequency of the electrons. That RM is "dark" (meaning invisible) and has virtually no absorption in the visible has been verified in numerous experiments. However, the situation is different in the infrared and RF ranges. In the IR, intracavity studies of RM (in the RM laser) show almost continuous stimulated emission from 800 to 16000 nm [Badiei and Holmlid 2003; Holmlid 2004a]. This strong optical activity probably extends through the far infrared region to radio frequencies [Holmlid 2007b] (see further below).

Manykin et al. calculated the radiative lifetime of RM in various excitation levels, and found it to be long, of the order of 100 years at an excitation level of $n_B = 16$. The main de-excitation processes for RM were found to be Auger processes [Manykin et al. 1992b]. The excitation level of RM in the laboratory is found to vary between the lowest possible value for the atom studied ($n_B = 3$ for Na and $n_B = 4$ for K) and an upper thermal value around $n_B = 40$ [Holmlid 2004a]. In comets and in interstellar space, the excitation level is generally $n_B = 80 - 100$ [Holmlid 2006a]. At $n_B = 40$ which may be a maximum value in the RM layer at the mesopause, the lifetime is extremely large (order of 1 Gy) from a simple extrapolation of this calculation. In the laboratory, the lifetime of RM with collisional quenching is of the range of minutes to several hours, depending on pressure and temperature. RM can be produced in various total pressure regimes. Methods to produce RM as clusters at low total pressure (down to $10^{-9}$ mbar) and low temperatures (from 500 K down to 10 K) have been developed [Wang and Holmlid 1998, 2000].

Since the Fermi level at $n_B = 40$ is at a distance of only 70 cm$^{-1}$ from the ionization level, ionization of such a highly excited RM cluster requires only a low energy photon, with a wavelength of the order of 150 μm. The ionization probability is normally at its maximum at the threshold, and falls rapidly at higher photon energies. Thus, the ionization probability due to visible photons will be quite small, a factor of $10^{-9}$ of the probability at threshold. Due to thermal electron emission over the very low work function barrier, RM will emit electrons which will give an electron density surrounding the RM layers [Badiei and Holmlid 2002b]. As observed in the laboratory, RM emits light in the IR (dark in the visible) and cools itself efficiently. Typical temperatures observed for RM clusters in a vacuum are below 20 K even at ambient temperatures of 300-600 K [Wang and Holmlid 2002; Badiei and Holmlid 2002c, d]. Thus, a thermal emission of electrons exists from RM clusters but at a lower rate than might be expected from the extremely low work function. The self-cooling to low temperatures is of course of interest for the cloud formation mechanisms discussed here.

The flux of fast particles and UV photons from the Sun will will not destroy RM but influence the excitation level $n_B$ of RM which will probably increase. Ground state atoms and molecules will become excited and also ionized. These ions recombine with electrons and form Rydberg species, which can be incorporated into the RM clusters. This gives a size increase of the clusters. Thus, the radiation is an important factor in the <u>formation</u> of RM, not in its destruction. The contact with ground state atoms and molecules in laboratory experiments mainly gives incorporation of the ground state species into the RM clusters with excitation energy sharing within the RM cluster. The direct addition of atoms to K(RM) clusters was studied experimentally by Wang, Engvall and Holmlid [1999]. Thus, collisions with slow particles do not destroy the RM clusters but collisions with ground state alkali atoms gives cluster growth.

RM clusters are most easily formed in desorption from carbon and metal oxide surfaces. Thus, carbonaceous and silicate dust particles in interplanetary space [Schulze et al. 1997] with their adsorbed and absorbed gas atoms and molecules provide an excellent direct desorption source for the formation of RM. The optical response of particles will be determined mainly by their electronic state. The polarizability of RM clusters is very large due to the almost free electrons in the RM. This can be observed from the strong Raman and stimulated Raman processes observed in RM [Olofson et al. 2003, Svensson and Holmlid 1999; Holmlid 2001]. This means that electronically excited species like RM clusters will dominate in optical studies over other particles like silicate and carbonaceous particles.

## 3. Results and discussion

### 3.1. Introduction to the Rydberg Matter interpretation

A novel analysis of the evidence from the features observed around the mesopause is given in the sections below. It is shown that the special and novel properties of RM which have been studied in the laboratory agree well with the observed properties of the material at the mesopause. The existing evidence is analyzed in the different subsections, starting in section 3.2 with the evidence concerning particle size and shapes from studies using linearly polarized light, circularly polarized light, and laser lidar methods. This section ends with a calculation of the height where the Na layers and thus the mesopause are found, directly given by the density of the RM clusters. In section 3.3, the measurements of the low temperature at the mesopause are described first. The relevant properties of the RM clusters concerning amplified spontaneous and stimulated emission in the visible and the infrared are described, and the very low temperatures in RM measured in the laboratory are given. After describing the properties of the RM clusters concerning excitation and deexcitation relevant for the seasonal variations of the temperature, the first accurate calculation ever is given of the lowest possible temperature at the mesopause, using the radiative properties of the RM clusters. In section 3.4, the radiative properties of the RM clusters in the microwave and radiofrequency regions are described from laboratory experiments, and these properties are shown to agree with the observed PMSE properties in radar heating and scattering. The present controversies in this field seem to be superseded by the RM description. The most spectacular properties of the mesopause namely the sporadic sodium layers discussed in section 3.5 fit accurately into the observed behaviour of RM in the laboratory, with Coulombic shock waves and condensation phenomena. The interesting observations of anti-correlation of Na layers with PMSE and NLC features also fits precisely with expectations from RM studies, as described in section 3.6. Finally, the very efficient formation of condensation nuclei for the final NLC formation through Rydberg reactions is briefly described in section 3.7, based on laboratory studies.

The basis of the RM description used here is in principle very simple. The main point is that RM layers exist at the mesopause as a natural consequence of the environment which exists there: low pressure, influx of sodium and other metals from interplanetary space, and the intense solar radiation especially during the summer. These RM layers are apparently the sources of all the different features observed, as summarized above and described in detail in the remainder of this paper. The RM description does not imply changes in the flux to or from the mesopause region relative to previous partial models, but it realizes that the alkali metals make a detour from the previously assumed relative uncomplicated path down to lower altitudes: instead, the alkali can stay for much longer times at the altitude of the mesopause in the condensed RM form floating at this air pressure and interacting with both the short-wave and the long-wave radiation passing through this layer.

The properties of RM are summarized above in section 2. However, it may be of interest to here briefly summarize the most important processes relevant to the RM clusters and layers and the evidence of them from laboratory experiments. (We only give a few references as examples). The RM clusters are proposed to be formed by alkali metal desorption from meteoritic particles, both carbonaceous and metal oxides (silicates). This RM desorption process has been studied extensively in the laboratory [Wang et al. 1999; Holmlid 1998b]. The RM formation process may take place during the initial interaction of meteorites with the atmosphere or later during slow descent as small particles. Alkali atoms in the alkali layer at the mesopause will become excited and form RM clusters (particles) during steady sunlit conditions (in the dark, they slowly de-excite and finally form ordinary metal particles). This is the reason why extensive RM layers mainly form close to the poles during summer. The excitation and de-excitation processes have been studied in the laboratory for example in optical cavities [Badiei and Holmlid 2005a; Holmlid 2007d]. RM clusters can also be transported to the mesopause from the surrounding interplanetary space, but if they come into the shadow, they will de-excite as well. The RM layers give the low temperature in this region by their strong optical activity at long wavelengths (see further below). The emission processes have been studied by spectroscopy in the laboratory, with [Holmlid 2004a; Holmlid 2008a] as examples. Since the RM clusters have high internal excitation energy, they can give Rydberg species and ions of most atoms and molecules in collisions with ground state and neutral species through energy transfer. In collisions with fast particles, they may also fragment to Rydberg species and ions themselves. Such processes have been studied by mass spectrometry and ionization methods in the laboratory [Lundin and Holmlid 1991; Wallin et al. 1993]. Thus a rich Rydberg and ion chemistry is started by the excess energy in the RM clusters: this excitation energy is originally derived from the sunlight. These two factors, the formation of condensation nuclei in the form of Rydberg species and ions and the low temperature, are the main factors giving condensation of water vapor. The water clusters give further particle growth and sedimentation of the ice particles, forming the visible NLCs.

This overview connects the various parts of a comprehensive description which has developed during a long time, for example the cometary and meteoritic material is now coupled to the low temperature in this region in a unique way. The RM layers have special properties in their radiative behaviour at long wavelengths, which gives the PMSE effects. This point is studied directly at radio frequencies in the laboratory [Holmlid 2007b, 2008a]. Our analysis mainly describes the inner working of the alkali layers, showing how the spectacular phenomena observed can take place and how they are linked together. The boundaries of our description agree with what is known otherwise about the alkali metal transport. Thus, our description in terms of RM features agrees with observations as well as most of the generally accepted partial models for the different phenomena. We may cite Höffner and Lübken [2007] stating "The upper mesosphere/lower thermosphere (UMLT) is probably the least understood region in the terrestrial atmosphere." The experimental studies cited above do not give all the rates for all the relevant processes, but the most significant ones have been quantized in a useful way in the literature. Thus, a complete numerical modelling of these very complex processes may be possible in the near future.

## 3.2. Formation of RM clusters and particles

At a considerable distance, RM clusters and particles are best observed by their infrared and radiofrequency emission patterns. This is used for their identification in the laboratory [Holmlid 2007b, 2004a] and in space, for example in comets [Holmlid 2006a] and in interstellar space in general as the so called UIR bands [Holmlid 2000]. Stimulated RF emission signatures like maser lines also indicate RM clusters in space [Holmlid 2006c]. In the upper atmosphere, such methods are probably not very useful due to the limited column density of material which exists there. This gives a low optical density against the IR background from interstellar or interplanetary space, showing the UIR-RM features [Holmlid 2000]. Light scattering methods are however also useful for RM identification, due the very large polarizability of RM clusters and the special symmetric and planar form of the RM clusters.

Linear polarization studies have been performed for a long time of NLCs and PMCs, using randomly polarized sunlight to excite the particles, and by observing the polarization state of the scattered light, also as a function of the angle in the scattering plane [Witt et al. 1976, Gumbel and Witt 1998]. Generally, the polarization is strong, which indicates Rayleigh scattering from small particles or molecules, and in most cases the observed behavior can be explained by spherical scatterers. Many different types of particles may give this type of scattering with a large degree of polarization of the scattered light. For example, small ice particles is one likely origin of this type of scattering. The existence of small water (ion) clusters is confirmed by sampling in situ with mass spectrometers in rocket flights (for example, Arnold and Viggiano [1982]). The most important information found is probably the particle size, usually found to be small and normally below 50 nm in diameter [Gumbel and Witt 1998, Gumbel et al. 2001]. This is of the same size as the most common RM clusters, as seen in Table 1. Pure RM clusters will give a relatively low degree of polarization, as observed in scattering from comets and asteroids [Holmlid 2006a, Levasseur-Regourd and Hadamcik 2003]. Thus it is unlikely that the linear polarization results are due only to RM clusters but to a large extent they are caused by water ice particles.

Circularly polarized light has been observed in scattering from NLCs in many experiments. The early experiments by Gadsden et al. [1979] are very significant from this respect. The possibility of prior polarization of the sunlight for example by multiple scattering has plagued such experiments on NLCs [Bohren 1983]. Modern lidar methods using linearly polarized laser light show a depolarization (cross polarization) of the backscattered light [Baumgarten et al. 2002], which can only be due to non-spherical scatterers. The layer of enhanced depolarization was located 1 km above the NLC layer, thus in the PMSE-RM range (see further below). Modelling used cylinder-shaped particles to match the observed result. Multi-wavelength lidar observations [Alpers et al. 2001] show a behaviour of the backscattering which can not be due to any of the simple shapes of the particles that were tested. The best agreement was found for Na-covered ice particles with 40 nm diameter, an unlikely construct however with some unique properties. Thus, it is apparent that particles with special properties must exist in the PMCs, and RM clusters is a very obvious alternative which has not been discussed previously in this context. RM clusters have been shown in experiments in the laboratory to have the properties required to explain the circular polarization features [Holmlid 2007c], and depolarization is also easily observed. In these experiments, linearly polarized visible light is scattered from the RM layer at an RM emitter surface, giving circularly polarized light with a strong left-right-handed asymmetry. See Fig. 3. Depolarization experiments are less informative, since many causes are possible. Thus, only the very significant RM circular polarization experiments are shown here. The main reason for the scattering of circularly polarized light from RM is that the conduction electrons in the planar cluster give a strong magnetic moment. The momentum of each orbiting electron in the plane adds to the other co-planar electrons, giving a large magnetic moment of each RM cluster equal to $m = Nn_B\mu_B$. Here, $N$ is the number of atoms in the cluster and $\mu_B$ is the Bohr magneton. This magnetic moment $m$ and the electron orbits provide a chiral entity which is the scatterer, giving circular scattering even for unaligned RM clusters. A circular polarization is also possible due to the optical properties of the RM phase. Due to the strong magnetic dipoles in the RM clusters, the clusters may also align partially in the geomagnetic field, especially when a low local temperature is created (see below). Such effects may often give the scattered visible light an even stronger circular polarization.

It is interesting that Alpers et al. [2001] find that Na coated ice spheres (initially due to Rapp and Lübken [1999]) with a diameter of 40 nm is a type of particle which agrees well with their multi-wavelength lidar experiments. This would indicate particles with high conductivity, very similar to the RM clusters which are metallic. However, the main reason that this type of particle agrees with their results is that the backscattering at their laser wavelength of 770 nm was very low, while the shorter wavelengths used were scattered much stronger. This phenomenon is more likely due to absorption at 770 nm. The observed decrease in backscattering intensity at the PMC altitude supports this explanation. In experiments in the laboratory, the infrared emission from RM is strong, giving broadband stimulated emission and even laser action. This means that absorption may also exist at these wavelengths. However, there is practically no emission from RM at wavelengths shorter than 700-800 nm, and only a few weak features exist in this visible light range [Holmlid 2007d]. In the visible wavelength range, the light will be scattered by the clusters with no absorption. It appears likely that the K D-line which is normally excited with this laser wavelength at 770 nm in the experiment was broadened since the K atoms were all embedded in (part of) the RM clusters, and the coupling to the RM clusters gave efficient absorption at the edge of the optically active region observed in the RM experiments. This effect was thus probably observed by Alpers et al. [2001] since very few free K atoms existed during their experiment.

The reason for the suggestion of Rapp and Lübken [1999] that Na covered ice particles existed in the PMCs was that they observed enhanced electron densities at these altitudes, and they suggested the only well-known system with low work function surfaces, i.e. alkali metal particles, to explain this. While alkali metals certainly have low work function surfaces, for example down to 2.3 eV for Na, it is also known that RM has the lowest work function of all materials, down to 0.3 eV [Svensson and Holmlid 1992]. This is so since RM is a low density metal, and a large interatomic distance results in a low work function according to metal theory [Lang and Kohn 1971]. Both experiments [Svensson et al. 1991] and theory [Manykin et al. 1992a, Svanberg and Holmlid 1994] confirm this, with experimental work functions down to 0.3 - 0.5 eV. This means that a phase with RM clusters will also have a large density of free electrons, of course varying with the temperature and the work function, in agreement with the findings by Rapp and Lübken [1999].

One important aspect of the PMCs is the time and place for their observation: at high latitudes during the local summer. This may mean that the low temperature and the cloud formation is initiated by sunlight. Of course, the visibility of ice clouds also requires that water vapour exists at high enough densities. In the RM description, the influence of sunlight is obvious, since the excitation and ionization of alkali atoms forming RM clusters and RM layers is driven by sunlight. These processes lead to Rydberg state formation through recombination of ions and electrons and RM cluster formation through desorption from the micrometeorite particles and further growth by additions of Rydberg atoms. Thus, the RM cluster formation depends directly on the solar radiation: in the shadow of the Earth, the clusters de-excite and finally form ordinary metal particles. That sunlight is responsible for the excitation is also apparent on Mercury and the Moon [Holmlid 2006b]. Sunlight will also promote aggregation and growth of the RM clusters and increase of the excitation level of the RM layers. This prolongs their life and the possibility of attachment of further alkali atoms to the RM layers by excitation energy transfer. Thus, it is expected that PMC clouds will be formed most easily during the local summer and that they will expand further, and definitely at high latitudes where the sunlight is continuous during this time of the year. During other seasons, the amount of RM formed from the alkali at this altitude will be smaller since sunlight is not available during such long periods. Of course, the density of water vapour may also be smaller for other reasons during other seasons. Thus, these very important aspects of NLC formation come out entirely natural in the RM description.

It is also interesting to attempt to derive the altitude of the RM layers, i.e the approximate altitude for the PMCs and PMSE. This is a quantity which is highly significant for the understanding of these phenomena. The density of an RM cluster or filament of RM clusters is roughly

$$D_{RM} = m / d^3 \quad,$$

where $m$ is the mass of an RM atom and $d$ is the distance between two atoms in the RM cluster. This distance is $2.9\, n_B^2\, a_0$, where $n_B$ is the excitation level and $a_0$ is the Bohr radius of 52.9 pm. This formula assumes that the distance between the clusters when they interact to form RM filaments [Badiei and Holmlid 2002a, b] is comparable to the value $d$ as shown in these references, or that the typical space taken for atoms in single clusters is comparable. The typical value of $n_B$ for interstellar RM is 40-80 [Holmlid 2000; 2007a]. We assume that $n_B =$ 40 is a more probable value for this low-temperature medium. This means a density of $3\times10^{-4}$ g dm$^{-3}$ for Na clusters. A slightly lower value of $n_B$ will give a larger density. The density of air at 90 km altitude and 120 K is approximately $6\times10^{-4}$ g dm$^{-3}$. Thus, the reason that the RM layers are at this altitude is that they have the same average density as the air, and Na will sink downwards only when the RM clusters become fragmented for example by collisions or de-excited by radiative processes.

**3.3. Low temperature**

Low temperatures, below 130 K, are observed in the metal containing altitude range around 90 km in numerous studies with many different techniques. Extensive observational work showing a deep minimum for the measured temperature at the mesopause has been reported. However, a calculation of the lowest temperature to be expected at the mesopause does not seem to exist (see further below). We may cite Höffner and Lübken [2007] stating "The upper mesosphere/lower thermosphere (UMLT) is probably the least understood region in the terrestrial atmosphere." The generally accepted upwelling and breaking of gravity waves and the resulting adiabatic cooling of air in the polar regions [Garcia and Solomon 1985] may give a partial and possibly transient description, but without a precise calculation of the resulting temperature the importance of such an effect is difficult to assess. Certainly, the well-known wave structure of the NLCs is likely due to breaking of waves from below. The non-adiabatic temperature gradients observed [for example Fritts et al. 2004] do not support this type of simple process. Further, the temperature minimum is stable even with varying slopes above and below the mesopause: this indicates a localized more permanent cause, not a turbulent one. The upwelling picture also seems hard to adapt to the formation of neutral alkali layers in the same region, since the alkali is due to flux from higher levels. Direct measurements of the vertical wind velocity indicate large turbulence at the mesopause [Fritts et al. 2004]. Thus the mesopause is at the borderline between the lower gravity wave influenced atmosphere which probably supplies the water vapour and the upper atmosphere from where the alkali metals are sinking down. The vertical wind and the eddy diffusion at the mesopause used as input by Raizada et al. [2007] contribute probably to the stabilization of and mixing in this boundary layer. It may finally be noted that the super-adiabatic temperature variation with altitude observed by Fritts et al. [2004] requires a radiative cooling process at the mesopause, as provided by RM and described below.

The best temperature measurements are made with lidar observations, for example at the K D-line [Eska et al. 1999], and by ionization gauge measurements on rocket payloads [Fritts at al. 2004]. Simultaneous measurements with different methods give good insight into the problems of the temperature measurements [see for example Williams et al. 2003, Fritts et al. 2004, Höffner and Lübken 2007]. (For further details about the methods used, see for example Von Zahn and Höffner [1996] for K lidar meaaurements and Rapp et al. [2004] for ionization gauge measurements). Sodium lidar measurements often give slightly higher temperatures than the other methods. Within the RM description, this may be due to the fact that the Na atoms observed by spectroscopy have been ejected from the RM clusters by energetic particle impact, and thus will have a higher temperature due to large initial kinetic energy. The release of K atoms from the RM clusters is likely to take place via another less violent mechanism. In agreement with this, a comparison of K lidar and Fe lidar temperature measurements indicates a lower temperature for the K atoms [Lautenbach and Höffner 2004].

RM clusters have been extensively studied in optical cavities, with thermal excitation of the clusters [Holmlid 2004a]. The emission from the cavity is due to stimulated emission from the electronically excited RM clusters, and the cavities have been operated as lasers [Badiei and Holmlid 2003] or as sources of amplified spontaneous emission (ASE) [Holmlid 2007d]. The strong emission in the infrared range indicates efficient emission and amplification, in one case with an observed gain of 100 m$^{-1}$ [Badiei and Holmlid 2003]. The detection of strong stimulated emission in the infrared from RM clusters was in fact preceded by the observation of very low temperatures of RM clusters by laser fragmentation. A nanosecond pulsed laser in the visible range induces Coulomb explosions in the RM phase by exciting electrons into higher localized states. The Coulomb repulsion is quantized and given by the excitation state $n_B$ of the RM. The flux of fragments from the explosions thus has a well defined kinetic energy, meaning that the fragments arrive at the detector within a very short time range, giving sharp time-of-flight peaks. One example is shown in Fig. 2. The width of the peaks is mainly due to the spread in translational velocities of the RM clusters prior to the fragmentation by the laser pulse, if the width of the laser beam focus (waist) is small enough relative to the distance to the detector. Many experiments on both $H_N$ and $K_N$ RM clusters indicate temperatures of 10-20 K [Wang and Holmlid 2002, Badiei and Holmlid 2002d]. This means that self-cooling due to stimulated emission in the almost collision-free environment with only RM clusters can reach very low temperatures. In Fig. 2, a further cooling effect exists due to RM formed by hydrogen, which means that the temperature of the peaks for light clusters is below 2 K.

In the metal containing layer at the mesopause, the density of Na is measured to be up to $4 \times 10^3$ cm$^{-3}$ in wintertime by satellite dayglow [Gumbel et al. 2007] or Na lidar measurements [She et al. 2000]. The K atom density is between 10 and 150 cm$^{-3}$, quite insignificant relative to Na [Eska et al. 1999]. The density of Na is a factor of ten higher in winter than in summer [Gardner et al. 1988]. We propose that the lower free Na density in the summer is due to increased excitation by sunlight and consecutive RM cluster formation, not to increased loss from the metallic layer as often assumed. In sporadic sodium layers (see further below), the Na density is found to be up to $10^5$ cm$^{-3}$. Thus, it can be concluded that the amount of Na bound in RM clusters in summer is at least ten times more than that observed as free atoms. The number density of RM clusters, assumed to typically be Na$_{19}$, is then of the order of 100 cm$^{-3}$ or higher. The pressure at this altitude is approximately $2\times10^{-3}$ mbar. Assuming that the geometric size of an RM cluster is its collisional cross section, the time between collisions with gas molecules is approximately 100 ns, very short for this pressure due to the large collision cross sections, while the time between collisions with other RM clusters is > 6 h.

The chemistry of the alkali metals in the upper atmosphere has been studied intensively by several groups. The removal of the alkali metals from the mesosphere has been of special interest, and the relatively slow processes of this type are not of direct concern to us here. We refer to studies especially by Plane and coworkers [Plane 2003]. The study of Fe, Na and K scavenging by water ice surfaces by Murray and Plane [2005] is very important in this connection. One special feature of the alkali metal densities that has received some attention but no explanation is the different seasonal variations of the Na and K densities. While the Na concentration is higher in the winter, due to less formation of the RM layers at low sunlight conditions as described above, the K atom concentration is a factor of 2-3 higher in summer than in the winter [She et al. 2006]. Using the RM description, this means that the K atoms are not so easily included in the RM clusters formed in the summer as the Na atoms are. In the winter, the K atoms are better accommodated in RM clusters. A closely related problem exists in the atmosphere of Mercury, where the density of K atoms is much lower than expected from several points of view, like the abundance in surface minerals and the corresponding atmospheric densities at the Moon. The same discrepancy is observed in the alkali layer at Earth, as discussed by Raizada et al. [2007]. The discrepancy at Mercury was explained by Holmlid [2006b] from two effects: the higher photon energy required to form Rydberg species and thus RM for Na due to its higher ionization energy, and the necessity of all atoms in an RM cluster to have the same excitation level. At the mesopause, both effects seem to contribute. The lowest excitation level of Na(RM) is $n_B = 3$, while K(RM) cannot be lower than $n_B = 4$. This excitation level may be reached after cooling at the bottom of the RM regions. This means that the larger K atoms in $n = 4$ are expelled from mixed Na,K(RM) clusters when the cluster are large and well organized in the summer. During the winter, K may form K(RM) clusters even during low-light conditions due to its lower ionization energy, experiencing less conflict with Na(RM) clusters during this period. Thus, the different seasonal variations of K and Na, and the too low density of K are all explained reasonably well within the RM description.

The cooling of the mesopause by the RM clusters will now be described, deriving a lowest possible temperature not including stimulated emission effects and transient wave phenomena. (The description is also valid for any particle with dominating absorption and emission processes at wavelengths much longer than the temperature maximum of the Earth). We know that the RM clusters radiate rapidly enough to cool down to very low temperatures, even if they still are in their electronically excited form. This is shown by the experiments in ultrahigh vacuum, giving values down to 2 K, as described above. Such values are observed even in a thermal radiation field from the surroundings at > 300 K, so the radiation density at an average active long infrared or RF wavelength is approximately 100 times larger from the clusters than from the surroundings at 300 K. This definitely indicates stimulated emission. The density of states that can take part in optical dipole (electric or magnetic) transitions is very large in the low wavelength range. There are not only electronic and rotational levels, as described above, but also broad vibrational bands (phonon bands) exist in the RM clusters [Holmlid 2008b]. When such clusters are located in a gas at low pressure, the temperature of the clusters will be determined also by the flux of energy to and from the colliding molecules, giving an average energy and thus temperature of the gas phase. Most molecules in the upper atmosphere do not interact strongly with long wavelength radiation since they are simple diatomic homonuclear molecules, but they interact via collisions with the RM clusters and may reach low temperature if the RM clusters radiate away their energy. The density of RM clusters is so low that stimulated emission is unlikely. However, the selective absorption and emission of the RM clusters at long wavelengths will still determine their interaction with radiation. The absorption and emission at long wavelength is described by the Rayleigh-Jean formula, which states that the emitted radiation power is proportional to the temperature. We have to realize that the atmosphere under such conditions would be at the surrounding temperature if the atmosphere was enclosed by a thermal bath at temperature $T$: consider the case of a perfectly reflecting mirror shell surrounding the atmosphere (microscopic reversibility). However, when radiation is leaking away the temperature will be lower, varying linearly with the total energy in the optically active long wavelength bands. The solid angle of the Earth seen by a point at 90 km altitude is approximately $1.67\pi$ sr, or a fraction of 0.418 of the total sphere. This means a local temperature of $0.418\,T$, and assuming an average surface (heat bath) temperature at Earth of 290 K in the summer means a lowest possible temperature in this layer of 121 K. This is close to what seems to be the lowest observed stable temperatures of 120 K [Fritts et al. 2004], for which the uncertainty may be close to 1 K. The mean temperature in summer given by Höffner and Lübken [2007] is as low as 119 K, which also agrees well with our calculation, assuming small variations in the ground temperature and the cited error limits of 1 - 5 K. If the temperature is lower at ground level, say 270 K as is possible in the Arctic region even in June, the temperature given by our calculational model would be as low as 113 K. Note that radiation from the Sun is not included in this calculation, since the Sun is often close to the horizon in measurements with NLC formation, and the effective absorption of IR radiation from the Sun in the mesopause layer is probably relatively small. Stimulated emission effects in RM give temperatures as low as 5 K in the laboratory [Wang and Holmlid 2002, Badiei and Holmlid 2002d]. If such effects exist at the mesopause, strong transient dips in the temperature could exist. This may be the reason for the unusually transient low temperatures down to 80 K measured by Lubken et al. [2009], but similar effects could also be due to gravity wave breaking phenomena. We conclude that the temperature at the mesopause is given quite accurately and probably uniquely by the calculational description using the selective absorption and emission features observed for RM clusters.

### 3.4. Radar scattering and heating (PMSE)

The polar mesospheric summer radar echos (PMSE) are often coupled to the PMCs by the time and altitude of their observations; and they are generally found to be at a slightly higher altitude than the PMCs but below the Na layer [She et al. 2006]. They are normally believed to be due to a so called dusty plasma, where the absorption and scattering of radio frequencies is modelled to be due to free electrons in the plasma. The dusty plasmas form well defined clouds at 80-90 km altitude [Biebricher et al. 2006]. The dust particles are of a size from approximately 50 nm with densities of $10^2$ cm$^{-3}$ (lidar measurements by von Cossart et al. [1999]), down to 10-20 nm with densities of more than $10^3$ cm$^{-3}$ for negatively charged particles [Havnes et al. 1996]. The sizes observed agree well with the sizes of RM clusters in Table 1. The densities agree quite well with the alkali densities and the Na RM cluster densities observed by lidar and satellites, as described more in detail in the preceding section.

Also, the increased electron density due to the small work function (ionization energy) of the RM clusters described above will simplify a description of the PMSE effects as dusty plasmas. The low work function of the RM clusters is probably the most important factor in the formation of negatively as well as positively charged particles reported by Havnes et al. [1996]. The characteristic features of the dusty plasma clouds are that they absorb, scatter and reflect VHF radar radiation, in the range from 50 MHz to above 1 GHz [Biebricher et al. 2006]. For example, the EISCAT radars are used at 224, 500 and 933 MHz for PMSE studies. These emitters are relatively broad band with 2-4 MHz bandwidth.

RM clusters are nearly metallic. This means that they will reflect radiation in most frequency ranges. This is apparently the reason for the strong radar echoes of the PMSE type. The electrons in RM are easily excited and driven by electric fields in their translational states in the metallic clusters. Thus, free electrons do not have to exist for reflection of RF radiation. However, the low work function of the RM clusters at high excitation levels means also that at thermal equilibrium, a large free electron density will exist. It may be expected that a dependence on the radar frequency as well as its intensity will exist for RM layers, since the absorption of radiation will heat the RM clusters and release free electrons. The discussion of the possible reflection processes by Rapp and Lübken (2004) indicates that high radar reflectivity requires a structure in the electron density in the medium. RM has an electron density variation due to the cluster form, with thermally emitted electrons as a cloud around each cluster. The large average distance between the clusters gives a further scale. Finally, the RM clusters at low temperature also form filaments as stacks of the planar clusters, as described by Badiei and Holmlid [2002a, 2002b] for interstellar space. This will however probably not be a dominating structure at the relatively high temperature of 120 K. These inhomogeneities will increase the scattering. The recent results by Belova et al. [2007] are very interesting for the reflectivity of PMSE. They found very different radar reflectivities for three radars simultaneously, at 53.5, 224 and 933 MHz. The reflectivites did not agree with a free electron model, and the authors concluded that no theoretical description could account for these results. The reflectivity was much lower at high (UHF) frequencies but these reflections came from higher altitudes. We propose this to be due to different sizes of the RM clusters at different heights, and possibly also to different excitation levels. The reflected spectral width was higher for the EISCAT-VHF than for the UHF radar, which could not be explained. We propose this effect to be due to a larger contribution by free electrons at the lower VHF frequency, and a larger contribution by bound RM electrons to the lower intensity scattering of the higher UHF frequency predominantly at high altitudes. Thus, these observations could be better explained by the RM description.

A quite different explanation for the PMSE layers has been proposed recently, where the scatterers are believed to be in the form of large (20-50 nm diameter) water ice particles, in a highly charged state [Rapp and Lübken 2004]. The discussion in this field has not concerned the reality of such water ice particles, neither of their charge state or of the amount of water vapour required for their almost constant formation, but has been more concerned with the possibility that such large water ice particles can give the RF scattering observed [Rapp et al. 2008]. If such particles would give the PMSE, it is difficult to see the difference between the PMSE and NLC clouds, which however are distinguishable features. PMSE and NLCs are even shown to occupy different altitudes and times [She et al. 2006], even if a development from PMSE to NLCs with time appears possible from their results. The RM description used here gives a better description of these phenomena.

We will further describe the agreement of the absorption (heating) RF features with the known properties of the RM clusters. In general, heating by RF radiation depends on absorption in relatively few and relatively narrow absorption bands and the question how the heating of the condensed RM clusters (relative to the heating of a plasma) can take place must be answered. RM clusters have large moments of inertia, and thus very small rotational energy constants. In the laboratory, stimulated emission radio frequency spectra are observed from such clusters, mainly studied from $K_N$ clusters. The frequency range studied so far is from 5 MHz up to 200 MHz. A typical experiment performed in a vacuum shows strong activity over large frequency regions in Fig. 4. The width of the lines is of the order of 10 kHz in this almost collision-free environment. The rotational constants can be easily calculated, and the rotational spectroscopy of the clusters involving several coupled angular momenta is well understood [Holmlid 2007b, 2008a]. The clusters are oblate symmetric rotors, and their energy levels follow standard formulas. The rotational constant in Hz is given by

$$B = \frac{h}{8\pi^2 I_b}$$

The moment of inertia that determines the pure rotational transitions is the moment $I_b$. This is the moment of inertia for rotation around an axis in the plane of the cluster. In the case of a symmetric planar cluster with $I_a = I_b$ due to symmetry, the moments of inertia are related as $I_c = 2I_b$. Here, $I_c$ is the moment of inertia for rotation around the figure axis (symmetry axis). The different moments of inertia $I_b$ for all planar six-fold symmetric clusters are easily calculated. They can be summarized for the most common clusters as $I_b/(m\,d^2) = 24$ ($N = 19$), 93 ($N = 37$), 255 ($N = 61$) and 570 ($N = 91$), where $m$ is the atomic mass, here in most cases the Na mass. This gives pure rotational transitions $\nu = 2B(J+1)$ for $\Delta J = \pm 1$, in fact $\nu \approx 2BJ$, since $J$ is large, of the order of several hundred at the temperatures of interest here. The spacings of the lines $2B$ for Na$_N$ clusters are easily calculated from the measured and theoretically predicted values for K$_N$ clusters [Holmlid 2007b, 2008a]. The value of $2B$ varies for example from 26.9 kHz for Na$_{91}$($n_B$ = 6) to 3.16 MHz for Na$_{19}$($n_B$=4). Na$_N$ clusters are also likely to form with $n_B$ = 3, with larger rotational constants. Absorptions and emissions take place at closely spaced frequencies in hundreds or thousands of lines for each cluster form, with a Doppler limited frequency width of a few tens of Hz due to their large mass. (The experimental linewidth for large RM clusters is often tens of kHz varying with the size of the cluster). The probablity of inclusion of K atoms in the Na clusters is quite high, even if the density of K atoms is much lower than the density of Na atoms. Such inclusions will form for example clusters KNa$_{18}$, with several different moments of inertia and rotational constants depending on how far from the center of the cluster the K atom is located. This increases the number of cluster forms strongly, contributing weak but numerous further lines. The rotational lines will be further broadened due to collisions. The time between collisions with gas molecules which may easily change the rotational energy quantum state was derived above to be approximately 100 ns. This will give a pressure broadening of the order of > 1 MHz. Thus, almost continuous absorptions in the RF range will exist for RM clusters from the kHz range upwards.

The rotational absorption will thus give broad rotational bands at RF frequencies. The broad maxima of these distributions will vary with the type of cluster and with the temperature. At 150 K, the maxima vary between 0.9 GHz for Na$_{61}$(5) and 5.5 GHz for Na$_{19}$(3). Thus, the RF range up to a several GHz will be densely covered with rotational transitions mainly from Na$_N$ clusters. This means that many different cluster types will become excited when the radar radiation reaches the PMSE (RM) layer. Absorption of the RF frequencies will heat the RM clusters from the absorptions described. The clusters will radiate mainly by spontaneous emission but maybe also by stimulated emission due to the long wavelength of the radiation: the average distance between the clusters is 1 mm, while the RF wavelengths employed are in the range 0.3 - 75 m. The stimulated emission in the RF range is discussed in Holmlid [2007b]. The maser lines observed in interstellar space also provide an excellent example of this type of property for the RM layers [Holmlid 2006c].

In experiments with the heating and VHF facilities of EISCAT, several interesting observations have been made, besides the strong interaction of the PMSE (RM) layers with the RF radiation discussed and rationalized above. The main effect studied by using the heater at 4-8 MHz is that the scattered signal at VHF frequencies is decreased when the heater is on. Such effects can be explained as due to heating of the electrons in a dusty plasma, which seems to be a fitting description of less well organized (hot) forms of RM [Norman 2001; Bonitz et al. 2004]. The RM description used here provides a larger density of electrons due to the low work function of the clusters, as described above. Heating of the RM layer may excite further electrons, and the dusty plasma description may be augmented by the greater availability of free electrons. For example, Biebricher et al. [2006] had to assume large, probably unrealistic densities of large 50 nm sized particles to find agreement with experiments. The RM description probably brings these values down to more realistic values. Other interesting features are observed in the form of so called overshoot and undershoot observations [Havnes et al. 2003; Biebricher et al. 2006]. This means that the VHF signal is observed to increase (or decrease) strongly immediately after the heater is turned off. The time constant for return to the steady-state scattering signal is approximately 30 s. This may be related to the time for recombination and formation of the RM layer after the thermal (RF driven) release of the electrons has ended. See below for a discussion of shocks in the RM layers, which is a phenomenon related to the RF heating.

## 3.5. Sporadic sodium layers and shockwaves

A short discussion is required to point out the most likely reason for the so called sporadic sodium layers. They have been observed by different methods, most significant by Na lidar measurements (for recent examples, we only mention Nesse et al. [2008] and Heinrich et al. [2008]). The sodium layers are observed to form thin (2 km) layers with a horizontal extension of 100 km to several 1000 km and a density of 2 – 20 times the background layer [Gu et al. 1995], i.e. up to $10^5$ cm$^{-3}$. They can be formed in a few minutes and exist for a few hours, disappearing rapidly or slowly. Since RM layers are invisible in many observational methods and certainly for the sodium lidar, RM layers with large dimensions can exist regularly at 90 km altitude. A strong disturbance like a collective ionization will give a Coulombic shockwave moving rapidly through the RM layer. This means that the collective state of RM is rapidly excited and disrupted, and is replaced by Na ions and electrons, forming atoms in excited Rydberg states and also partly in the ground state by rapid recombination. The real density of Na is thus probably much higher than the ground state observations with lidar give information of. The collective ionization process may be initiated spontaneously or by an external event like energetic particle impact: this is possibly the reason for the relation between auroral phenomena and sporadic sodium layers as reported by for example Gu et al. [1995].

A shockwave in RM formed of H atoms was studied in Badiei and Holmlid [2005b] while shocks in RM formed from potassium were studied by Åkesson et al. [2006]. This type of shock is a Coulombic shock, propagating by field ionization of neighbouring Rydberg species. In the case of hydrogen, the shock velocity was measured to be close to $2\times10^4$ m s$^{-1}$.

This means that such a shock will move 1000 km in 50 s. Thus the shocked RM visible to the sodium lidar will appear in a time of the order of a few minutes even over distances of the order of 1000 km, in agreement with observations. No other satisfactory mechanism for this part of the process seems to have been proposed. Such events require of course that large RM layers exist with a high enough density: we propose that the time of storage of the sodium in the form of RM at high altitude is long and thus the density high. At too high air densities the shock will be attenuated, delayed and finally stopped. The disappearance of the sodium layer is due to renewed condensation of the excited sodium atoms to RM clusters. This process may be slow or fast depending on the excitation state and density of the sodium atoms. The suggested RM mechanism thus seems to meet all the requirements summarized by Nesse et al. [2008]. Finally, it should be noted that similar almost planet-wide de-excitation (ionization) events giving short-duration resonance fluorescence from the sodium clouds in the atmosphere at Mercury are well known [Killen et al. 2007]. The low density sodium atmosphere at Mercury is concluded by Holmlid [2006b] to contain RM clusters that act as temporary storage in the cycling of Na between the surface and the upper atmosphere of Mercury, with excitation by radiation from the Sun maintaining the density of RM clusters.

### 3.6. Anti-correlation of features

In a recent study of NLC and PMSE features and of the simultaneous Na density, She et al. [2006] report that the density of Na decreases where PMSE or NLCs are formed. In general, the NLCs are observed at altitudes lower than the PMSE, which is at lower altitudes than the Na layers. The anti-correlation reported is spatial in character, such that the formation of PMSE is observed at lower altitudes than the main Na density. The anti-correlation is strong relative to NLCs, but somewhat weaker relative to PMSE. The authors conclude that the Na atoms become bound to small particles in the NLCs and PMSE events, thus decreasing the amount of observable Na atoms. That alkali atoms are easily bound to water ice with a large accommodation coefficient was shown directly in laboratory experiments by Murray and Plane [2005], which may be one reason for the decrease in Na density where NLCs are formed. This interaction has been reported previously. That a similar phenomenon also exists when PMSE is formed is a new proposal by these authors.

From a general and logical point of view, the anti-correlation of Na with PMSE and NLCs should first be tested against possible explanations of all these features as due to various forms of Na, not introducing the concept of water ice at all until it is absolutely needed and positively identified, as is done far down in the case of NLC. The main proposal to describe the anti-correlation features should logically be based on the various forms of Na, atomic at large altitudes and condensed as RM at lower altitudes. In the RM description, the reason for the decrease in Na density in both cases is the formation of Na(RM) clusters and clouds, not the attachment to water ice particles. At large altitudes, Na atoms exist and become excited by sunlight to Rydberg states. These condense and form a RM and PMSE reflective layer, which sinks due to its relatively high density and the low temperature of RM. At the bottom of this PMSE layer, the boundary to the upwards directed flux of air with a larger density of water vapour exists. There, the RM layers decompose due to collisions and form ions (see further below), which give condensation nuclei and the large ice particles visible as NLCs. This type of sequence is expected and agrees with observations. This description does not require that water vapour rises so high in the atmosphere as the PMSE, up to 90 km according to She et al., but instead freezes out in the NLC clouds at somewhat lower altitudes. This makes the RM description better than the ice particle model for the PMSE, since it seems unlikely that water vapour can reach such large altitudes. To distinguish between the RM and water-ice particles, the density of water vapour should be measured above the PMSE layer. With the RM description, a complete and consistent picture of the alkali-RM-PMSE-NLC features is obtained.

### 3.7. Formation of condensation nuclei

The rates of chemical reactions vary rapidly with the reaction cross sections of the reacting species, being of the order of 0.01 nm$^2$ for neutral – neutral reactions and 0.1-1 nm$^2$ for ion – neutral reactions. This means that ion – neutral reactions dominate over neutral – neutral reactions, especially at low temperatures since the cross sections for such processes even increase at low velocities (temperatures). Rydberg species (circular states) at principal quantum number $n = 40$ have a radius of the electron orbit of 80 nm, which means a cross section for collision of 20 000 nm$^2$. Such enormous cross sections are probably only valid for energy transfer collisions, and Rydberg <u>reactive</u> cross sections are considerably lower. However, it is still likely that Rydberg reactions, both Rydberg – Rydberg and Rydberg – neutral reactions take place first, after a release of free Rydberg states. Such reactions will give ions, which will then take part in ion – neutral reactions. Thus, the most effective start of chemical reactions is of course the release of Rydberg species in high excitation levels, because of their high reactivity and the large electronic excitation energy they carry. Due to the periodic (sporadic sodium layers) and also more or less continuous breakdown of RM clusters into Rydberg species, the RM layers will give an energy input to many reaction chains, forming ions of many different types. Thus, ion chemistry is expected to be initiated by the RM layers, likely also in the end giving condensation nuclei for ice particle formation.

Rydberg reactions are notoriously difficult to investigate in the laboratory by laser methods, at least if state resolution (specification of the initial electronic state) is intended. This is so since laser methods are incapable of forming the most important (the circular) Rydberg species: to form a $n = 40$, $l = 39$ atomic state, 39 consecutive laser pulses with different wavelengths would be needed, with almost no atoms reaching the final state intended. Distributions of Rydberg species over many relatively large quantum numbers can however be formed easily by various forms of recombination experiments, and such experiments have been performed.

The excitation energy of an alkali Rydberg atom (maximum 5.1 eV for Na) is not large enough to ionize a molecule like $N_2$ (15.6 eV), which is the most common molecule in the altitude range of interest here. This point has been studied for example by Wallin et al. [1993]. Thermally excited Rydberg atoms of Cs were found to form ions like $N_2^+$, and it was concluded that this process was possible since doubly excited states (or reactive RM clusters) were formed in the experiments. Earlier studies with other methods gave reaction cross sections of 40 – 100 $nm^2$, clearly indicating Rydberg reactions by the large cross sections. Recent spectroscopic results [Holmlid 2007d] show that doubly excited alkali states K** are formed by thermal processes, and other studies give direct proof that doubly excited states of alkali atoms are formed when RM is de-excited [Kotarba and Holmlid 2008]. The final degradation of the initial electronic excitation energy in alkali atoms A* may be shown with a few (oversimplified) reaction steps as

$$A^* + A^* \rightarrow A^{**} + A \quad \text{(energy pooling)}$$

$$A^{**} + M \rightarrow A^* + M^* \quad \text{(energy transfer)}$$

$$A^* + M \rightarrow A + M^+ + e^- \quad \text{(ionization)}$$

$$M^* + M \rightarrow M + M^+ + e^- \quad \text{(ionization)}$$

where A** is a doubly excited alkali atom. In such states, the excitation energy is high enough to give excitation and ionization of most molecules, as verified by experiments. The ionization energy of $H_2O$ is 12.6 eV, thus lower than for other molecules like $N_2$ which gives ions in such processes in the laboratory. It may be of interest that all electronically excited states of the water molecule have a Rydberg character, which probably simplifies the transfer of excitation energy from Rydberg species to this molecule.

We here point at the likely origin of excitation energy providing large ion densities in the mesosphere. This couples together the different aspects of the PMSE-PMC-nucleation processes. The ion chemistry for which we only describe the initiation steps in a new way has been extensively studied in the laboratory. We refer the interested reader to the gas-kinetics and atmospheric literature, for example to the textbook by Wayne [2000].

## 4. Conclusions

Based on extensive laboratory experiments, we suggest that the distinctly different features studied at an altitude of approximately 90 km in the atmosphere (at the summer mesopause) all may be explained by a common suitable candidate material, namely layers of the excited condensed material Rydberg Matter (RM). These layers are composed mainly of Na atoms, but also contain K and possibly H atoms. Sporadic sodium layers and PMSE give direct evidence for the existence of RM layers at the mesopause. Since RM is a condensed material, spectroscopy can not observe any Na atoms in the RM phase. The origin of this material is partially interplanetary space, where similar clouds exist in comets, at Mercury and the Moon. RM is also formed by solar excitation of the alkali brought to the upper atmosphere in meteoritic form. The altitude of the RM layers is correctly calculated from the density of the RM, and the layers can not sink further down since their density is the same as air at this level. The average summer temperature of the RM layers is calculated from the intense optical activity at long wavelengths to be down to 121 K at this altitude, which seems to be the first calculation ever of this value. The reason for the increased activity of PMSE, sporadic sodium layers, PMCs and NLCs during the summer is probably the increased radiation during this period, which stabilizes the RM clusters and layers: They probably also increase in size and excitation level due to the radiation, as observed in the laboratory. Most of the sodium density is believed to be locked up in the RM layers in the summer: this means that a strong increase of free sodium density is observed during the winter when the RM layers decrease in density due to the smaller solar radiation. Rydberg initiated chemistry is faster and more efficient than ion chemistry and gives ions even of molecules with large ionization energy, in the end providing condensation nuclei for PMC formation. Special aspects like the difference between the seasonal variations of K and Na atoms are tentatively described on the basis of the RM description, in agreement with observations from other planetary atmospheres.

Table 1.

Values of the experimental or theoretical (in parenthesis) interatomic distance $d$ (in nm) in Rydberg Matter clusters, with excitation level $n_B$ and magic number $N$. The experimental values are for clusters $K_N$. The second value in each case is the calculated (experimental or theoretical) diameter of the cluster.

| $n_B =$<br>$N =$ | 4 | 5 | 6 | 7 | 8 |
|---|---|---|---|---|---|
| 19 | 2.4105 ± 0.0002<br>9.6 | 3.808 ± 0.001<br>15.2 | 5.525 ± 0.002<br>22.1 | (7.52)<br>30.1 | (9.82)<br>39.3 |
| 37 | (2.46)<br>14.7 | (3.84)<br>23.0 | 5.46 ± 0.02<br>32.8 | 7.520 ± 0.008<br>45.1 | 9.82 ± 0.02<br>58.9 |
| 61 | (2.46)<br>19.6 | 3.749 ± 0.005<br>30.0 | (5.53)<br>44.2 | (7.52)<br>60.2 | (9.82)<br>78.6 |
| 91 | (2.46)<br>24.6 | 3.764 ± 0.005<br>37.6 | 5.364 ± 0.008<br>53.6 | (7.52)<br>75.2 | (9.82)<br>98.2 |

**Figure captions**

Fig. 1. A cluster of Rydberg Matter with 19 atoms. This type of 6-fold symmetric cluster is one of the main types found in experiments on Rydberg Matter clusters. Its diameter is 15.2 nm at excitation level $n_B = 5$.

Fig. 2. Several forms of $K_N$ RM clusters observed by time-of-flight measurements from pulsed laser induced Coulomb explosions in an RM cloud. The kinetic temperature of the first peaks is below 2 K, likely due both to stimulated emission cooling and collisional cooling by hydrogen gas admitted in this experiment. The excitation level for each cluster type is given in parentheses. In the upper panel, the experimental arrangement inside a vacuum chamber is shown. Note that the laser beam passes <u>outside</u> the central part of the RM emitter (observe the perspective in the drawing). Similar results are published in for example Åkesson et al. [2006], Holmlid [2006b] and Badiei and Holmlid [2002c, d].

Fig. 3. Circular polarization observations in backscattering from an RM surface layer on a metal oxide surface used as RM emitter. The top panel shows the strong left-right asymmetry of the scattering from the RM layer, while the lower panel shows the apparent spectra observed by a grating spectrometer [Holmlid 2004c].

Fig. 4. The strong activity in the radio frequency range of RM clusters due to rotational transitions of the clusters. The data are observed with a bandwidth of 3 kHz in emission from RM clusters. The strong peaks are due to $K_{19}$ in excitation level $n_B = 4$ with different J numbers, while the smaller peaks are due to $K_{91}$ in $n_B = 5$. The clusters are located in a vacuum chamber with an RM emitter at a few hundred K, as shown in the upper panel. Similar results can be found in Holmlid [2007b, 2008a].

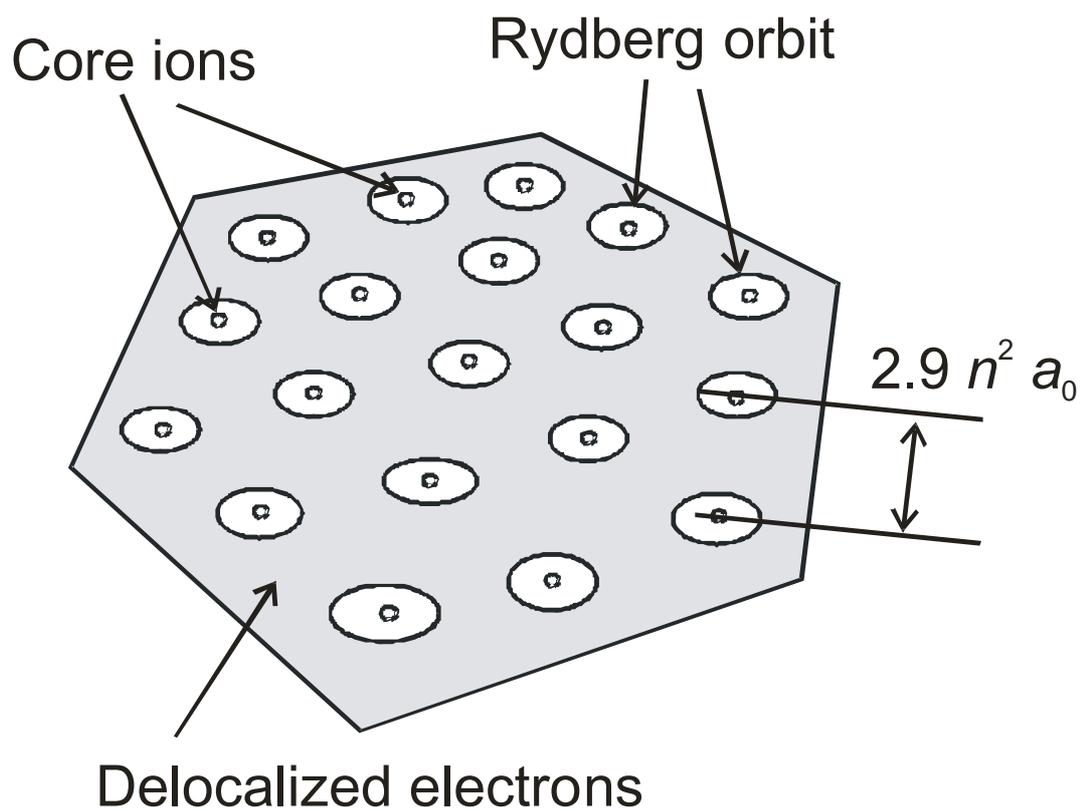

Fig. 1

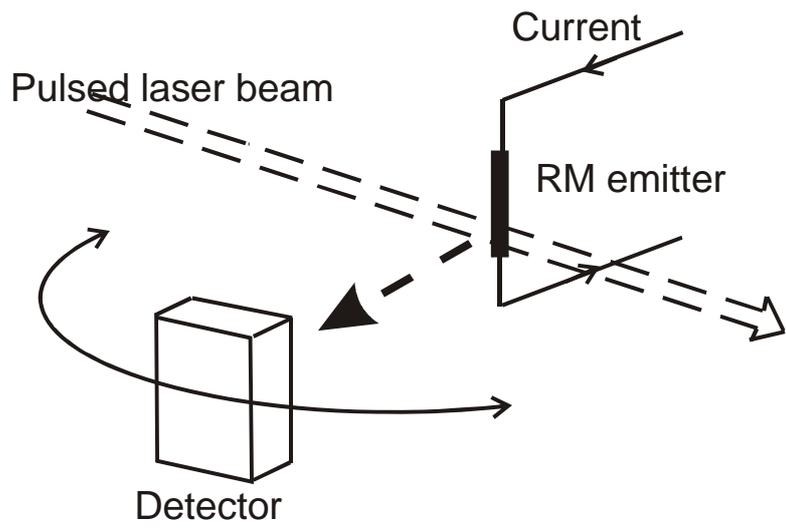

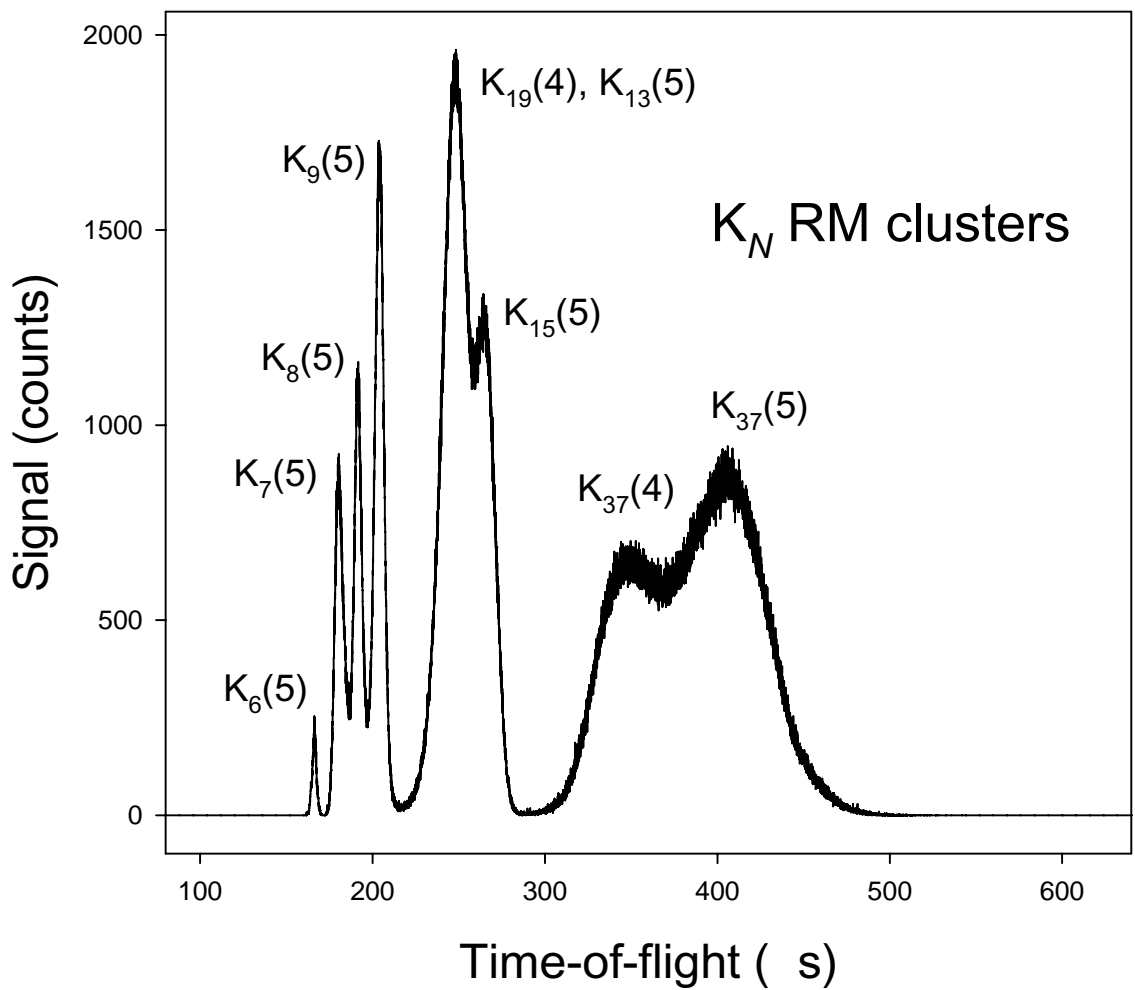

Fig. 2

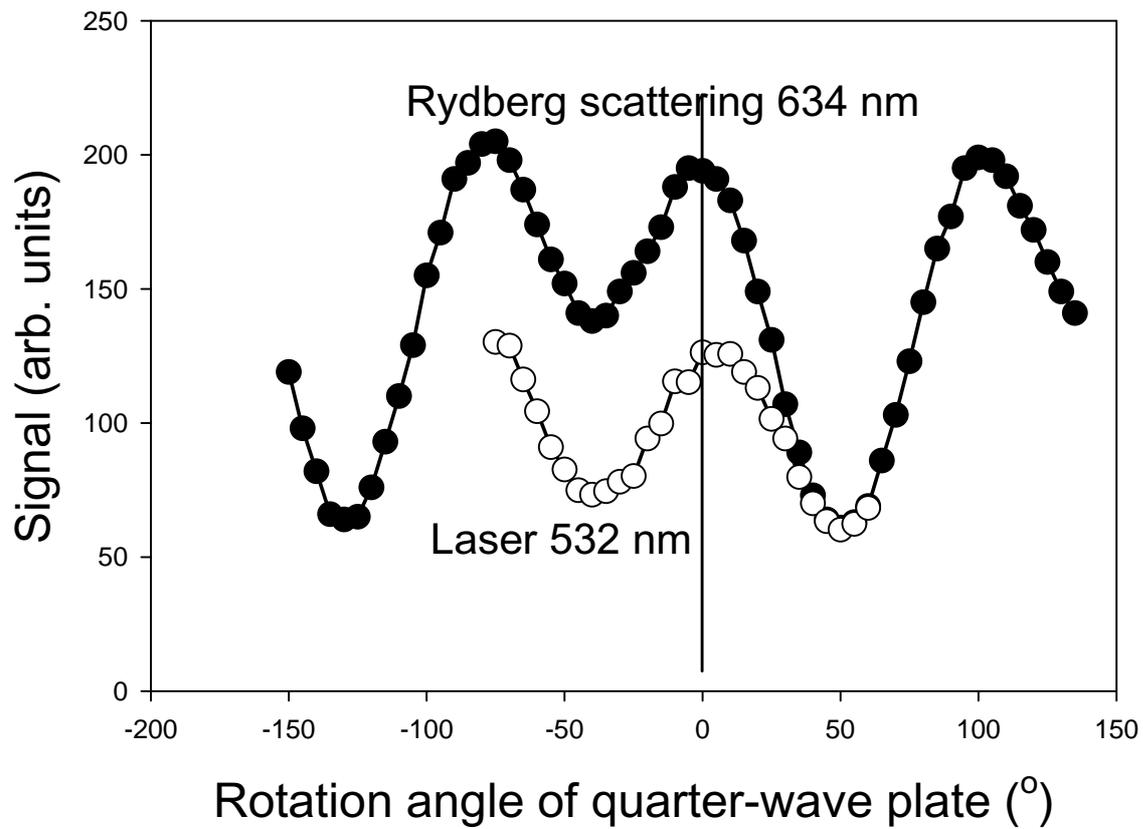
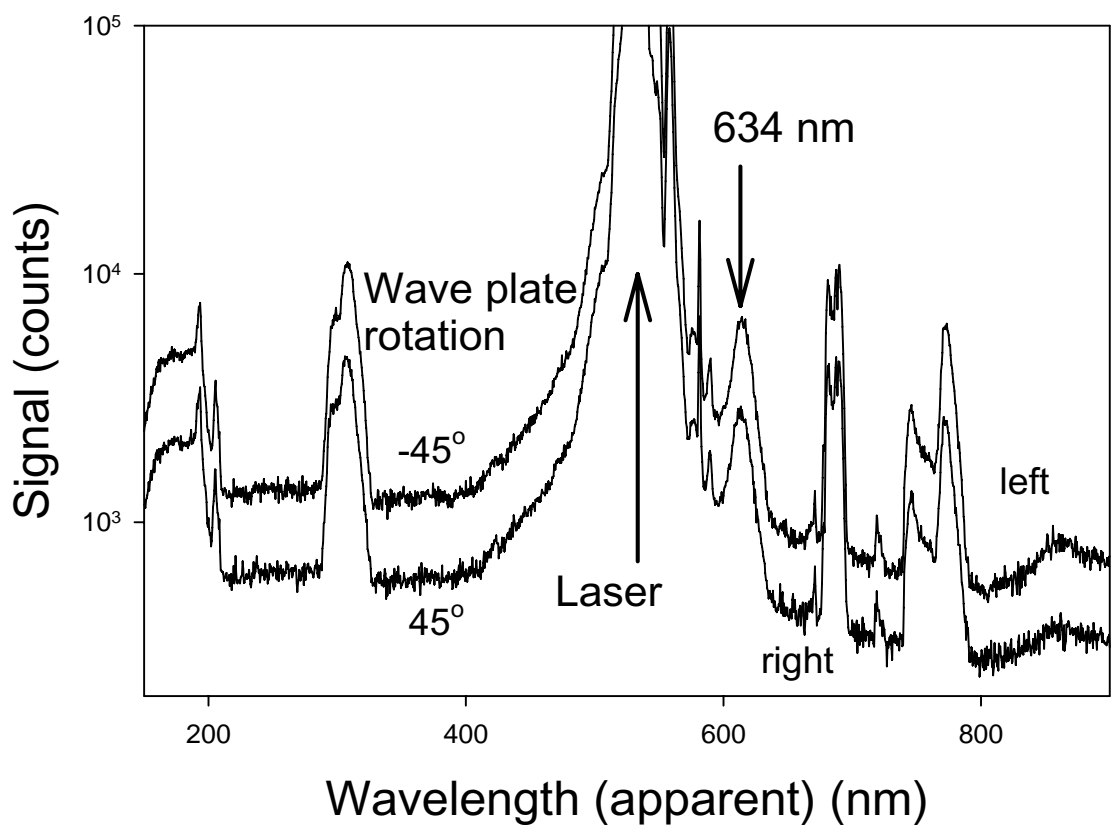

Fig. 3

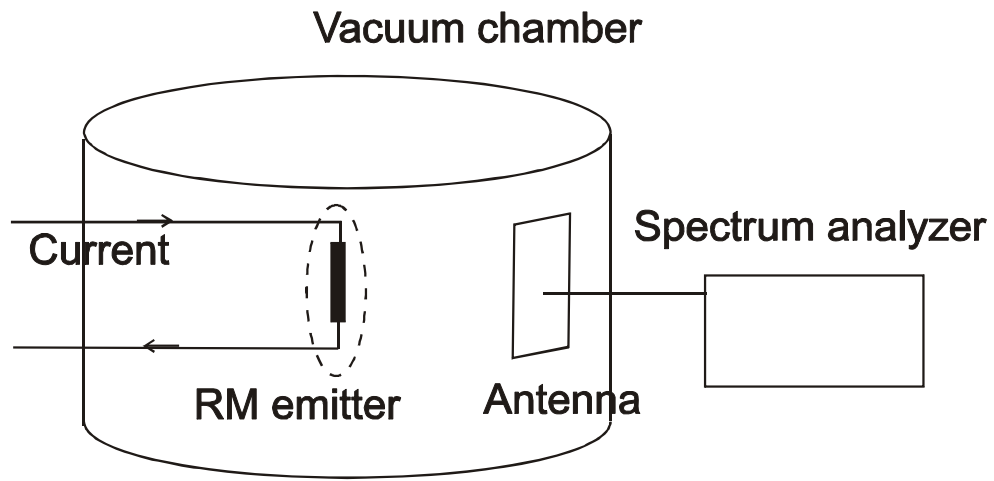

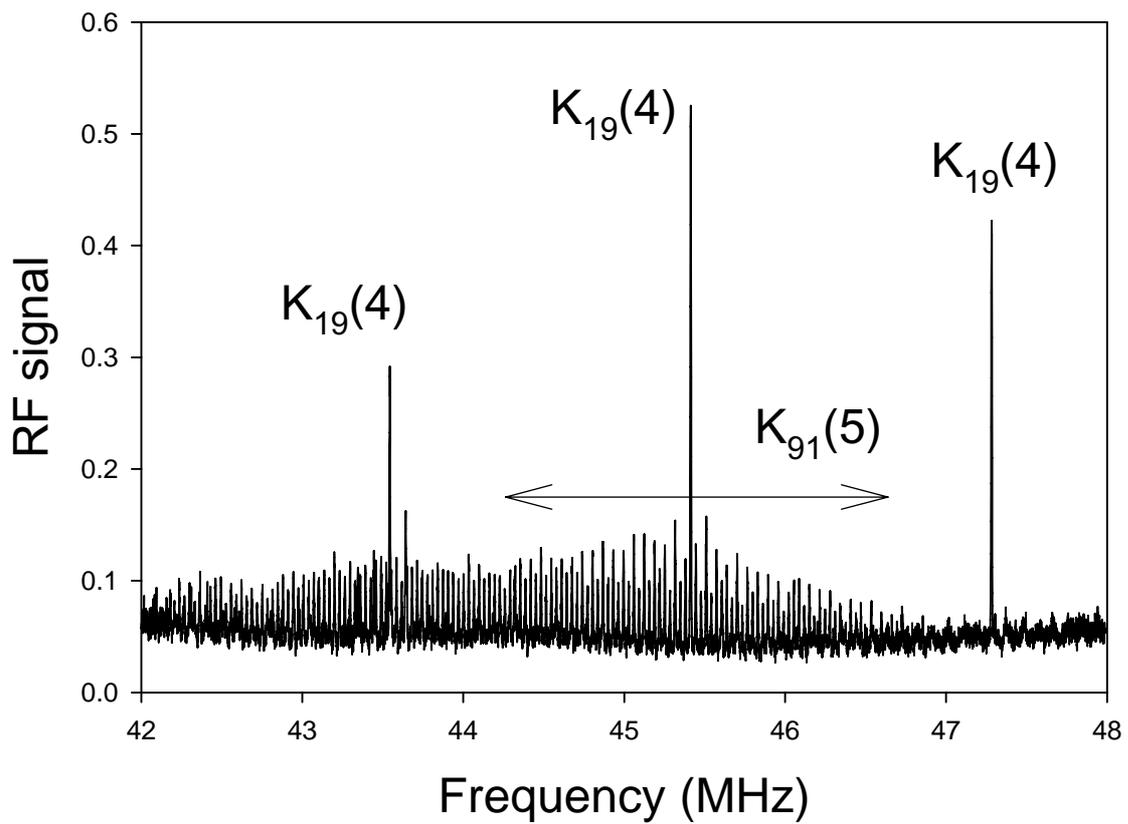

Fig. 4